\documentclass[12pt]{article}
\usepackage{lscape, amsmath, amssymb, amscd}

\usepackage{epsf}
\usepackage{ifthen}
\usepackage{times}

\newcommand{\thd}[1]
{\ifthenelse {\equal{#1}{1}}
	{{#1}$^{\mathrm{st}}$} 
	{{\ifthenelse{\equal{#1}{2}}{{#1}$^{\mathrm{nd}}$}{{#1}$^{\mathrm{th}}$}}}}

\usepackage{amsmath}
\usepackage{amsthm}

\begin{document}
\renewcommand{\textfraction}{0}
\newtheorem{thm}{Theorem}

\title{Performance of PPM Multipath Synchronization in the Limit of Large Bandwidth  }
\author{\normalsize Dana Porrat \\
\small School of Engineering and Computer Science \\
[-5pt] \small The Hebrew University \\
[-5pt] \small Jerusalem, Israel \\
[-5pt] \small dporrat@cs.huji.ac.il
\and
Urbashi Mitra \\
\small Department of Electrical Engineering\\
[-5pt] \small University of Southern California \\
[-5pt] \small Los Angeles, CA USA \\
[-5pt] \small ubli@ucs.edu
}

\date{}
\maketitle
\thispagestyle{empty}

\begin{abstract}
The acquisition, or synchronization, of the multipath profile for
an ultrawideband pulse position modulation (PPM) communication
systems is considered. Synchronization is critical for the proper
operation of PPM based For the multipath channel, it is assumed
that channel gains are known, but path delays are unknown. In the
limit of large bandwidth, $W$, it is assumed that the number of
paths, $L$, grows.  The delay spread of the channel, $M$, is
proportional to the bandwidth.  The rate of growth of $L$ versus
$M$ determines whether synchronization can occur.  It is shown
that if $\frac{L}{\sqrt{M}} \rightarrow 0$, then the maximum
likelihood synchronizer cannot acquire any of the paths and
alternatively if $\frac{L}{{M}} \rightarrow 0$, the maximum
likelihood synchronizer is guaranteed to miss at least one path.

\end{abstract}

\normalsize

\section*{Acknowledgement}
The authors thank Alex Samorodnitsky for his participation and help on statistical matters.
\section{Introduction}

We consider the asymptotic performance of maximum likelihood
synchronization schemes for pulse position modulations (PPM) in
the limit of large bandwidth. This study is motivated by recent
interest in ultra wideband (UWB) signaling schemes for radio
communications. While significant diversity is achievable given
the large amount of multipath in UWB channels, methods for
harnessing such diversity when the channel is unknown remain a
challenge. In fact, the fundamental limits of UWB signaling in the
presence of channel uncertainty have not been fully established.
In this work, we focus on the uncertainty in determining the
multipath profile (including leading delay) of the UWB channel.
For the purposes of this work, such uncertainty is equivalent to
uncertainty in synchronization. In particular, knowledge of
channel path delays is critical for the operation of pulse
position modulation systems over large bandwidths.  Timing errors
as small as fractions of nanoseconds can seriously degrade  system
performance as reported in \cite{Lovelace02,Tian}. Furthermore,
\cite{VWA04} suggests that threshold-based ultra wideband (UWB)
synchronization for PPM does not perform well even in
asymptotically high SNR; further implying that the performance of
pragmatic synchronization could limit their UWB potential.  We
have shown in \cite{dana_ubli_isit} that in the limit of large
bandwidth, threshold based synchronizers cannot achieve
synchronization.

For fixed bandwidth spread spectrum systems (see \cite{CM98} and
references therein), it has been observed that detectors are more
sensitive to mismatch in delay information versus other channel
parameters. Information theoretic analysis of spread-spectrum
systems\cite{channel_uncertainty} shows that the scenario of
unknown path gains with known delay locations achieves almost the
same throughput as that of complete channel knowledge (gains,
delays) in the limit of large bandwidth.  In contrast, for unknown
channel parameters, the throughput diminishes  in the limit of
large bandwidth if the number of channel paths increase faster
than a logarithmic rate on the bandwidth.

This work shows that pulse position modulation is very sensitive
to the knowledge of path delays. In fact, even a mild increase
(with bandwidth) of the number of identically and independently
distributed paths composing a channel will cause a PPM system to
fail in the limit of large bandwidth. The rate of increase
dictates system performance. This paper is organized as follows:
Section~\ref{sec-signal}
 describes the transmitted signal, channel model and received
signal. The main result for maximum likelihood synchronization is
provided in Section~\ref{sec-ml}.  Discussion of the new result
with respect to our prior work is given in
Section~\ref{sec-prior}.

\section{Signal Model}
\label{sec-signal}
\subsection{The Transmitted Signal}

We consider pulse position modulation (PPM), where the transmitted
signal can be written as
\begin{eqnarray*}
x(t) &  = & \sum_{n=-\infty}^{\infty} p \left(t - nT_s - \frac{1}{W} b[n]\right)\\
p(t) & = &  \left\{  \begin{array} {c c} \sqrt{ \frac{\mathcal{E}}{\theta}}  & t\in \left[0, \frac{T_s}{N} \right) \\
0 & \mbox{else} \\ \end{array} \right .
\end{eqnarray*}
The symbol duration is given by $T_s$ and the number of pulse
positions is dictated by the transmission bandwidth $W$, {\em
i.e.} $N= W T_s$.
The data symbol is denoted $b[n] \in \{0, 1,
\cdots, N-1\}$.
$\mathcal{E}$ is the average transmitted energy per symbol, that is bandwidth independent, and $\theta$ is a flash parameter to be explained shortly.
Thus, in a symbol duration, there is a single rectangular pulse of duration $\frac{T_s}{N}$.
Our goal is to investigate performance of such a PPM system as the transmission bandwidth increases.
We shall assume that the symbol duration does not diminish; however, due the use of flash signaling \cite{verdu_2002}, the information rate will not grow without bound.
With the use of flash signaling, transmission is bursty and communication occurs over a fraction $\theta$ of the total communication period.
The {\em flash parameter} $\theta$ is known at the receiver and furthermore, the receiver is aware of the on-periods of communication.
The transmission frame corresponds
to a coherence period of the channel and as such, one out of every
$\frac{1}{\theta}$ coherence periods is employed for transmission
(an on-period).

Note the distinction between flashy transmission and PPM modulation.
For regular data transmission,
the receiver must detect which one of the $N=WT_s$ pulse positions has been employed in each symbol; in contrast, with flashy transmission, the receiver is synchronized to the on-periods of communication.
We note that if $\theta$ is quite small, then the transmitter is predominantly silent.

The fraction of time utilized for transmission may decrease as the
bandwidth $W$ increases, but it cannot do so too fast.
In order to
maintain a positive (non-diminishing) data-rate, the parameter
$\theta$ must be large enough so that $\theta\log W$ does not
diminish.
The reasoning for this is straightforward:  $\log_2 W
T_s$ bits are transmitted per symbol; however, only a
fraction of the coherence periods are employed and thus the data rate is proportional to $\theta \log_2 W T_s$.
The requirement on $\theta$ can be written as:
\begin{eqnarray}
\theta\geq \frac{k_1}{\log \left( Wk_2 \right)} \label{eq:theta}
\end{eqnarray}
with fixed $k_1,\ k_2$ that are independent of the bandwidth.
\par
Several features of our setup should be underscored. The first is
that there is no limit imposed on the number of PPM positions that
are employed for data signaling. Thus, a guard time can be
implemented by limiting the positions employed. Second, we
emphasize the employment of a lower bounded symbol time, where the
lower bound does not depend on the signal bandwidth. We do not
consider schemes where the symbol time diminishes with bandwidth.
Thus, the number of bits that can be transmitted in a single
coherence period depends logarithmically on the bandwidth.
Note that systems that use a guard period between symbols, that depends on the channel path delays, have a natural lower bound on their symbol time.

\subsection{The Channel and Received Signal}

We assume an tapped delay line model for the
channel $h(t)$, thus
\begin{eqnarray*}
h(t) & = & \sum_{l=1}^L g_l \delta \left(t - \frac{d_l}{W} \right)
\end{eqnarray*}
where the channel gains are given by $g_l$ and we assume
that $\sum g_l^2 = 1$; $\delta(\cdot)$ denotes the Kronecker delta
function and $d_i$ represent the path delays which are assumed non-negative integers between 1 and $M$.
For simplicity of exposition we shall assume a uniform profile for the path gains and therefore
$g_l = \frac{1}{\sqrt{L}} \; l=1,\dots,L$.
The maximal possible number of
resolvable paths is given by $M=W T_d$, where $T_d$ represents the
maximum delay of the channel, thus the actual number of paths $L$ must satisfy $L \leq M$.
Recent wideband
channel propagation measurements suggest that the number of
channel paths grows sub-linearly with bandwidth \cite{rusch_2002},
possibly satisfying
\begin{eqnarray*}
\lim_{ W \rightarrow \infty} L =  \infty & \mbox{and} & \lim_{ W
\rightarrow \infty} \frac{L}{W} = 0
\end{eqnarray*}
Given $M$ possible values of the path delays, we assume that the
realizations of the path delays are uniformly distributed over ${M
\choose L} = \frac{M!}{L! (M-L)!}$ possibilities.
The channel
model is of the block-type:  the channel is fixed over the
channel coherence time $T_c$;  channel realizations at different
coherence periods are statistically independent.
\par
The received signal is given by,
\begin{eqnarray*}
y(t) & = & h(t) \otimes x(t) + z(t) =\sum_{l=1}^Lg_l
x\left(t-\frac{d_l}{W} \right)+z(t),
\end{eqnarray*}
where $z(t)$ is a zero-mean, white Gaussian noise process.

At the receiver, the received signal is matched filtered with the
pulse shape and sampled at $\frac{1}{W}$ yielding the following
discrete time equivalent signal:
\begin{eqnarray}
Y_i &= &\frac{1}{\sqrt{L}}\sum_{l=1}^L X_{i-d_l}+Z_i \label{eq:discrete}\\
X_i & = & \left\{\begin{array}{ll}
\sqrt{ \frac{ \mathcal{E}}{\theta}}   &
\mbox{if }\exists n:i \div N = n \\
& \mbox{and  } i\;\mathrm{mod}\; N=b[n] \\
0 & \mathrm{else}
\end{array}\right. \nonumber
\end{eqnarray}
$i\div N$ signifies the largest integer $k$ such that $kN\leq i$.
The signal $X_i$ is zero-valued except at the positions
corresponding to the transmitted PPM pulse; recall that $N=W T_s$
is the number of possible positions.
The amplitude of the signal
at the non-zero position is normalized so the noise samples
$\left\{Z_i\right\}$ are zero-mean with unit variance.

In order to assess the challenges of synchronization of PPM in
multipath,  we analyze a further simplified system that operates
under two additional conditions. We assume a sufficiently large
guard time,$T_d$, to ensure {\em no intersysmbol interference}
resulting in an effectively longer coherence time $
\tilde{T_c}=\frac{T_s+T_d}{T_s}T_c $.  And we assume {\em
knowledge of the PPM symbols}, essentially assuming training
information. The receiver sums over all the symbols per coherence
period before it begins processing. Given that we show the failure
to synchronize for an optimal detector under these idealized
conditions, we effectively make statements about more practical
systems as well.

\section{Maximum Likelihood Synchronization}
\label{sec-ml}

Recall that the position of the PPM symbol is known; however, the
initial delay and multipath profile are unknown.  The
synchronization problem can be posed as a multiple hypothesis
testing problem for which there are $M \choose L$ hypotheses given
a delay spread of $M = W T_d$ and $L$ non-zero channel taps.  The
received signal under each hypothesis can be written as, (recall
(\ref{eq:discrete})),

\begin{eqnarray}
\underline{Y}|H_i & = & \left[ Y_0, Y_{1}, \cdots Y_{M-1}
\right]^T =\sqrt{\frac{ \mathcal{E} }{ L \theta}}
\underline{s}_i + \underline{Z} \\
\underline{s}_i & = & \left[ \underbrace{1, 1, 0, \cdots 1 ,0
,}_{\mbox{$L$ 1's over $M$ positions}} \right]^T \; \; \; \; \;
\underline{Z} \sim  {\cal N}\left(\underline{0}, {\bf I} \right)
\end{eqnarray}
The optimal detector for such a scenario is a simple correlator:
\begin{eqnarray}
\hat{i} & = & \arg \max_i \underline{s}_i^T \underline{Y}
\end{eqnarray}
However, given the form of $\underline{s}_i$, we can see the
following equivalence.  Let $\tilde{Y}_1, \tilde{Y}_2, \cdots,
\tilde{Y}_L$ be the $L$ largest components of $\underline{Y}$.
Then,
\begin{eqnarray}
 \max \underline{s}_i^T \underline{Y} = \sum_{j=1}^L  \tilde{Y}_j
\end{eqnarray}
Thus the maximum likelihood detector is equivalent to determining
the multipath locations by selecting the $L$ positions with the
$L$ largest signal values.  With this perspective of the maximum
likelihood detector, we can develop a method for evaluating the
likelihood of an error through order statistics.  The statistics
of signal positions and noise positions are Gaussian and given by,

\begin{eqnarray}
Y_i | \mbox{path location} & \sim & {\cal N} \left( \sqrt{
\frac{\mathcal{E}}{ \theta L}}, 1 \right) \label{eq:sig}\\
Y_i | \mbox{noise only position} & \sim & {\cal N} \left( 0, 1
\right) \nonumber
\end{eqnarray}
Given that we know the PPM symbol, the observation vector is of
length $M$ and of the $M$ possible positions, $L$ correspond to
the transmitted signal.
The remaining $M-L$ correspond to
noise.

We present the main theorem of the work:

\begin{thm} Consider $M$ independent, Gaussian random variables with the
following distributions,

\begin{eqnarray}
Y_i & \sim & {\cal N}\left(\sqrt{k\frac{\log M}{L}},1\right), \ i
= 1, 2, \cdots, L\\
W_i &\sim & {\cal N}(0,1), \ i = 1, 2, \cdots, M-L
\end{eqnarray}
$k$ is a constant which does does not depend on $L$ or $M$; $L
\leq M$. We order the $Y_i$ such that $B_1 = \max Y_i$ and $B_L =
\min Y_i$; similarly, $S_1 = \max W_i$ and $S_{M-L} = \min W_i$.
 Then,
 \begin{eqnarray}
\lim_{M,L \rightarrow \infty} P \left[S_L > B_1 \right] =1 & &
\mbox{if} \; \;
\frac{L}{\sqrt{M}} \rightarrow 0\\
\lim_{M,L \rightarrow \infty} P \left[S_1 > B_L \right] =1& &
\mbox{if} \; \; \frac{L}{{M}} \rightarrow 0
\end{eqnarray}
\end{thm}

\par  If $\frac{L}{\sqrt{M}} \rightarrow 0$, then the maximum
likelihood detector will always detect noise variables and none of
the correct paths will be detected.  On the other hand if a faster
growth rate on the multipath exists, $\frac{L}{{M}} \rightarrow
0$, the maximum likelihood detector is guaranteed to miss at least
one of the path locations.  In the limit of large $L$, missing one
path is insignificant; however, we have two limits on bounds on
performance.

\par
 To provide intuition about our result consider a detector that
 {\bf randomly} selects positions, where the selection does not depend on the amplitudes.  If we assume a uniform
 selection of $L$ variables out of $M$, this selection will include, on average, $L^2/M$ of the $L$ big variables.
To see this, look at an example with $L=1/2M$. The probability
that each big variable is chosen is $1/2$, so on average the
number of big variables chosen is $1/2L=L^2/M$. With $L\geq
c\sqrt{M}$, the average number of big variables chosen this way
does not diminish as $M$ increases. The optimal detector performs
better than a random choice, so it detects at least $L^2/M$ of the
$L$ big variables. This is the reason for the condition
$L/\sqrt{M}\rightarrow0$.

\subsection{Outline of Proof}

The proof follows from a simple observation of the events leading
to an error for the maximum likelihood detector.  That is,
$P_e^{ML} = P \left[ \cup_i A_i \right]$, where the $A_i$ are
error events where one or more noise positions are members of the
set of $L$ largest values.  Thus, $P_e^{ML} \geq P[A_i]$ for any
$i$.  We consider two particular error events:  one path is
incorrectly detected and all paths are correctly detected.  In the
context of the ordered random variables  we can view these events
as:  $P_e^{ML} \geq \max_i P[A_i]$ which corresponds to a single
path error -- this is the most likely error to occur and $P_e^{ML}
\geq \min_i P[A_i]$ which corresponds to all paths being
incorrectly detected -- this is the least likely error to occur.
However, for both events, we shall show that for reasonable growth
rates on $L$ versus $M$, the probability of these two events is
unity in the limits of large $L$ and $M$.  As $M$ is proportional
to the bandwidth of the system, we ultimately obtain a large
bandwidth result.  We shall consider order statistics on the
signal and noise variables.  We shall show that the mean of the
relevant noise variable dominates over the mean of the relevant
signal variable.  For $\frac{L}{\sqrt{M}} \rightarrow 0$, the
maximum likelihood detector finds none of the signal positions,
leading us to the conclusion that under these conditions, no
detector can synchronize.

\subsection{Useful Formulas}

From {\em Cram\'er}'s book~\cite{cramer_book}, Section 28.6, page
376, we have the mean and variance of the order statistics of
Gaussian variables. We order $G$ identically and independently
distributed Gaussians with mean $m$ and variance $\sigma^2$. The
$\nu^{\mathrm{th}}$ variable from the top has mean
\begin{equation}
E_{\nu:G}=m+\sigma\left(\sqrt{2\ln G}- \frac{\ln\ln G+\ln 4\pi+2\left(S_1(\nu)-C\right)}{2\sqrt{2\ln G}}+O\left(\frac{1}{\ln G}\right)\right) \label{eq:mean_o}
\end{equation}
and variance
\begin{equation}
\mathrm{var}_{\nu:G}= \frac{\sigma^2}{2\ln G}  \left(\frac{\pi^2}{6}-S_2(\nu)\right)+O\left(\frac{1}{\ln^2G}\right) \label{eq:var_o}
\end{equation}

where
$C\approx0.5772$ is Euler's constant, and for $\nu>1$
\begin{eqnarray}
S_1(\nu) & = & \frac{1}{1}+\frac{1}{2}+\dots+\frac{1}{\nu-1} \\
S_2(\nu) & = & \frac{1}{1^2}+\frac{1}{2^2}+\dots+\frac{1}{\left(\nu-1\right)^2}
\end{eqnarray}
For $\nu=1$ (the highest of the $G$ variables), we have from~\cite{johnson_book}, Section 21-4, page 278:
\begin{eqnarray}
S_1(1) & = & S_2(1)= 0
\end{eqnarray}

\subsection{The $L^\mathrm{th}$ Largest Noise Variable}

We show that the $L^\mathrm{th}$ largest of the noise variables
equals $\sqrt{2\ln \left(M-L\right)}- \frac{\ln L}{\sqrt{2\ln
\left(M-L\right)}}$ in the limit of large $M$, $L$. We
use~(\ref{eq:mean_o}) and~(\ref{eq:var_o}) with $G=M-L$ random
variables, and consider  the $\nu=L$ largest variable, {\em i.e.}
there are $L-1$ variables larger than the one we investigate. We
recall that $m=0$ and $\sigma=1$, which yields the following mean
for the variable of interest,
\begin{equation}
E_{L:M-L}=\sqrt{2\ln \left(M-L\right)}- \frac{\ln\ln \left(M-L\right)+\ln 4\pi+2\left(S_1(L)-C\right)}{2\sqrt{2\ln \left(M-L\right)}}+O\left(\frac{1}{\ln \left(M-L\right)}\right)
\end{equation}
and variance
\begin{equation}
\mathrm{var}_{L:M-L}=\frac{1}{2\ln \left(M-L\right)}\left(\frac{\pi^2}{6}-S_2(L)\right)+O\left(\frac{1}{\ln^2\left(M-L\right)}\right)
\end{equation}
where
$C\approx0.5772$ and
\begin{eqnarray}
S_1(L) & = & \frac{1}{1}+\frac{1}{2}+\dots+\frac{1}{L-1} \\
S_2(L) & = & \frac{1}{1^2}+\frac{1}{2^2}+\dots+\frac{1}{\left(L-1\right)^2}
\end{eqnarray}
We observe that $S_2(L)$ is finite for any $L$,  thus $\lim{L,M
\rightarrow \infty} \mathrm{var}_{L:M-L} = 0$.  Therefore, in the
limit of large $M$ and $L$, the $L^\mathrm{th}$ largest variable
approaches a constant which is equivalent to the mean.

To further investigate the mean,  we calculate a simple approximation for
$S_1(L)$, for large L:
\begin{eqnarray}
S_1(L) & = & \frac{1}{1}+\frac{1}{2}+\dots+\frac{1}{L-1}\approx
\int_1^L \frac{1}{x}dx = \ln x|_{1}^L=\ln L
\end{eqnarray}
\begin{eqnarray}
E_{L:M-L} & \approx& \sqrt{2\ln \left(M-L\right)}- \frac{\ln\ln \left(M-L\right)+\ln 4\pi+2\left(\ln L-C\right)}{2\sqrt{2\ln \left(M-L\right)}}+O\left(\frac{1}{\ln \left(M-L\right)}\right) \\
& \approx & \sqrt{2\ln \left(M-L\right)}- \frac{\ln L}{\sqrt{2\ln
\left(M-L\right)}}
\end{eqnarray}

Observe that $\lim_{M,L \rightarrow \infty} E_{L:M-L} = \infty$,
furthermore, it is straightforward to show that
$\sum_{M=L}^{\infty} \mathrm{var}_{L:M-L} = \infty$.  Thus, the
$L^\mathrm{th}$ largest variable does not converge to a limit in
the mean square sense (see {\em e.g. }\cite{Viniotis}) despite the
fact that $\lim_{M,L \rightarrow \infty} \mathrm{var}_{L:M-L} =
0$.  Thus, we can only conclude that the random sequence of
interest converges in distribution.

\subsection{The Largest Signal Variable}

We show that the largest of the signal variables approaches its
mean value,  $\sqrt{k\frac{\log M}{L}}+\sqrt{2\ln L}$, in the
limit of large $M$, $L$, if $L^2<M$.  We use
Equations~(\ref{eq:mean_o}) and~(\ref{eq:var_o}) again, with $G=L$
random variables, and examine the largest variable, that is
$\nu=1$.  Recall that $m=\sqrt{k\frac{\log M}{L}}$ and $\sigma=1$,
yielding the mean,
\begin{eqnarray}
E_{1:L} & = & \sqrt{k\frac{\log M}{L}}+\sqrt{2\ln L}- \frac{\ln\ln L+\ln 4\pi-2C}{2\sqrt{2\ln L}}+O\left(\frac{1}{\ln L}\right) \\
& > & \sqrt{k\frac{\log M}{L}}+\sqrt{2\ln L}
\end{eqnarray}
and variance
\begin{equation}
\mathrm{var}_{1:L}= \frac{1}{2\ln L}  \frac{\pi^2}{6}+O\left(\frac{1}{\ln^2L}\right)
\end{equation}
where $C\approx0.5772$. As in the previous case,
 $\lim_{M,L \rightarrow \infty} E_{1:L} = \infty$, and
$\lim_{M,L \rightarrow \infty} \mathrm{var}_{1:L} = 0$.

\subsection{Conditions for Dominance}

We seek to determine the conditions for which $ E_{L:M-L}>
E_{1:L}$.  Thus, the desired strict inequality is given below,
\begin{eqnarray}
\sqrt{2\ln \left(M-L\right)}- \frac{\ln L}{\sqrt{2\ln \left(M-L\right)}} & \gg  & \sqrt{k\frac{\log M}{L}}+\sqrt{2\ln L}  \\
\sqrt{2\ln \left(M-L\right)}& \gg  & \sqrt{k\frac{\log
M}{L}}+\sqrt{2\ln L}  + \frac{\ln L}{\sqrt{2\ln \left(M-L\right)}}
\label{eq:es}
\end{eqnarray}
To assess this comparison, we compare each of the three terms on
the right hand side of~(\ref{eq:es}) with three terms of a
decomposition of the left hand side. That is, let,
\begin{eqnarray}
\sqrt{2\ln \left(M-L\right)} & = &
\left(\alpha+\beta+\gamma\right)\sqrt{2\ln \left(M-L\right)} \; \;
\; \mbox{where} \; \; \alpha+\beta+\gamma=1
\end{eqnarray}
$\alpha$, $\beta$ and $\gamma$ are positive constants which are
not functions of $M$ and $L$.  We next execute three comparisons.
For the first term of~(\ref{eq:es}), we square both sides of the
inequality of interest to achieve,
\begin{eqnarray}
\alpha^22\ln \left(M-L\right)  & \gg? &  k\frac{\log M}{L} \\
\alpha^22\left(\ln M+\ln\left(1-\frac{L}{M}\right)\right) & \gg? &
k\frac{\log M}{L}
\end{eqnarray}
The last equation is an inequality for any positive $\alpha$,
$L,M\rightarrow\infty$ and $\frac{L}{M}\rightarrow0$. Consider the
second term of~(\ref{eq:es}),we also square both sides and yield,
\begin{eqnarray}
\beta^2\ln \left(M-L\right) & \gg ? & \ln L \\
\left(M-L\right)^{\beta^2} & \gg ? & L \\
M^{\beta^2}\left(1-\frac{L}{M}\right)^{\beta^2} & \gg ? & L \\
M^{\beta^2}& \gg ? & L
\end{eqnarray}
The final equation is an equality as long as $\beta^2>0.5$, due to
the constraint of small $L$, that is $L^2 < M$. Examining the
final term of~(\ref{eq:es}):
\begin{eqnarray}
\gamma\sqrt{2\ln \left(M-L\right)} & \gg?&  \frac{\ln L}{\sqrt{2\ln \left(M-L\right)}}  \\
\gamma2\ln \left(M-L\right) & \gg?&  \ln L \\
\left(M-L\right)^{2\gamma} & \gg?&  L \\
M^{2\gamma} & \gg?&  L
\end{eqnarray}
We achieve in equality when $2\gamma>\frac{1}{2}$, again due to
the relationship between $M$ and $L$.

Summarizing the conditions, we need to determine a set of
$\alpha,\beta$ and $\gamma$ such that, (a) $\alpha+\beta+\gamma=1$
(b)$\alpha$ is any positive constant (c)
$\beta>\sqrt{\frac{1}{2}}$ and (d) $\gamma> \frac{1}{4}$. However,
these conditions can always be met, thus for the appropriate
growth rates on $L$ and $M$ ($\frac{L}{\sqrt{M}} \rightarrow 0$),
the mean of the $L^\mathrm{th}$ largest noise variable dominates
the mean of the largest signal variable.

We assume the required conditions above.  We form a new random
variable $D_{M,L} = S_L-B_1$.  As the signal and noise variables
are independent, we can determine that ${\bf E} \left[D_{M,L}
\right] =  E_{L:M-L}- E_{1:L}$ and ${\bf var} \left[D_{M,L}
\right] = \mathrm{var}_{L:M-L}+\mathrm{var}_{1:L}$.  Furthermore,
$\lim_{M,L \rightarrow \infty}
 {\bf E} \left[D_{M,L}\right] = \infty$ and $\lim_{M,L \rightarrow \infty}
 {\bf var} \left[D_{M,L} \right]= 0$.

  From the Chebyshev inequality we can show
  \begin{eqnarray}
\rightarrow \; \; \lim_{M,L \rightarrow \infty} P \left[ |D_{M,L}
- {\bf E} \left[D_{M,L}\right]| <\epsilon \right] &\geq& 1 -
\lim_{M,L \rightarrow \infty} \frac{ {\bf var} \left[D_{M,L}
\right]}{\epsilon^2} = 1
\end{eqnarray}
  In the limit of large $M$
and $L$, this difference variable approaches its mean with
probability 1; recall that this mean value is infinity. Thus, the
limiting distribution of the difference variable, $D_{M,L}$ is a
delta function.  Using Fatou's theorem which enables the
interchange of limits and integration, we determine that
$\lim_{M,L \rightarrow \infty} P \left[D_{M,L}
>0 \right] = \lim_{M,L \rightarrow \infty} P \left[S_L
> B_1 \right] =1$.

The proof for the comparison of $S_L$ and $B_1$ is similar, we
note that the key statistics/approximations are given as follows
for the largest noise variable:
\begin{eqnarray}
E_{1:M-L}& \gtrsim &\sqrt{2 \ln (M-L)}\\
 \mathrm{var}_{1:M-L} & =
&\frac{\pi}{12\ln
\left(M-L\right)}+O\left(\frac{1}{\ln^2\left(M-L\right)}\right)
\end{eqnarray}
and for the smallest signal variable,
\begin{eqnarray}
E_{L:L}
 & \approx& \sqrt{k\frac{\log
M}{L}}+\sqrt{\frac{\ln L}{2}}\\
\mathrm{var}_{L:L} & = & \frac{1}{2\ln L} \left( \frac{\pi^2}{6}-
S_2(L) \right)+O\left(\frac{1}{\ln^2L}\right)
\end{eqnarray}

Thus, the above arguments prove the theorem of the work.  We note
that the corollary statement is that if the growth rate of the
number of paths, $L$ versus the delay spread $M$ is such that
$\frac{L^2}{M} \rightarrow 0$, then the maximum likelihood
multipath detector cannot synchronize any path and if $\frac{L}{M}
\rightarrow 0$, at least one path will be incorrectly detected.
Two features of practical UWB channels and systems have not been
considered herein, but are under current investigation.  The first
is that we have bounds on the relative growth rates of multipath
to delay spread for the two extreme cases:  no paths synchronized
and all but one path synchronized.  Of  interest is to determine
what ratio of paths is necessary.  Finally, the multipath profile
considered is for equal energy paths.  In reality, the path energy
appears to decay from the first path.

\section{Relationship to Prior Results}
\label{sec-prior}
 In \cite{dana_ubli_isit}, we provided two
results. The first was that there did not exist a threshold for
which position by position threshold detection could achieve
arbitrarily small probability of error for multipath
synchronization in the limit of large bandwidth. The second result
determined the conditions under which maximum likelihood
synchronization {\bf could} achieve an arbitrarily small
probability of error in the limit of large bandwidth.  The
conditions for the maximum likelihood detector from
\cite{dana_ubli_isit} are:

\begin{eqnarray*}
L & < & \sqrt{\frac{ \mathcal{E} \log (W k_2)}{ k_3  (\log 2) W
T_d}}
\end{eqnarray*}
where, $k_2,k_3$ are constants which are independent of $L$ and
$W$.  Recall that $W$ is the bandwidth and $T_d$ is the delay
spread.  As $M$ is proportional to the bandwidth, the conditions
above imply  that we want the behavior of $L \sim o\left(
\sqrt{\frac{ \log M}{M}} \right)$\footnote{This is {\em little-o
notation}. That is, $f(n) = o(g(n))$ if $\exists k \ni, f(n) < c
g(n) \; \; \forall \; n \geq k$.}.  Thus if the rate of growth on
$L$ is as just noted, we have $\lim_{M \rightarrow \infty} L
\approx \lim_{M \rightarrow \infty} \sqrt{\frac{ \log M}{M}} =0 $.
Thus, the required condition is that the number of multipath
actually diminish with increasing bandwidth.  This is equivalent
to the energy of the multipath profile concentrating itself in
proportionally fewer and fewer components.  Recall that the mean
of the unordered signal variables is given by $\sqrt{k\frac{\log
M}{L}}$, thus as there are fewer and fewer paths, the energy of
the non-zero paths increases making them more ``detectable''.  We
contrast this result to result of the current work.  Herein, if
$\lim_{M,L \rightarrow \infty} \frac{L}{\sqrt{M}} = 0$, the
optimal detector cannot synchronize. In this case, the number of
paths does grow without bound and thus the energy in each non-zero
path is decreasing to zero, simultaneously, the number of noise
positions is dominating the number of signal positions. As such,
the mean of the larger noise positions is increasing, while the
mean of the smallest signal position is in fact decreasing,
leading to the large likelihood of an error.

\section{Conclusions}
\label{sec-conc} In this work, we have considered the problem of
channel synchronization for PPM modulation for ultrawideband
communication systems.  Even under idealized conditions of no
intersymbol interference and perfect knowledge of transmitted
symbols and channel gains, the optimal synchronizer fails when the
rate of growth of the number of multipath grows too slowly
relative the bandwidth.  We have shown that for
$\frac{L}{\sqrt{M}} \rightarrow 0$, the maximum likelihood
synchronizer fails to capture any of the paths and for
$\frac{L}{{M}} \rightarrow 0$, the maximum likelihood synchronizer
is guaranteed to miss at least one path.  Ongoing research is
considering the effect of the profile of the gains -- equal energy
paths are not consistent with experimental channel data as well as
the necessary growth rates when a fraction of the multiple paths
need to be properly synchronized.

\bibliography{refs}

\begin{thebibliography}{10}

\bibitem{CM98}
L.-C. Chu and U.~Mitra.
\newblock Performance analysis of the improved {MMSE} multi-user receiver for
  mismatched delay channels.
\newblock {\em IEEE Trans. on Comm.}, 46(10):1369--1380, October 1998.

\bibitem{cramer_book}
Harald Cram\'er.
\newblock {\em Mathematical Methods of Statistics}.
\newblock Princeton University Press, 1946.

\bibitem{johnson_book}
Norman~L. Johnson and Samuel Kotz.
\newblock {\em Continuous Univariate Distributions - 1}.
\newblock Houghton Mifflin Company, 1970.

\bibitem{Lovelace02}
W.~M. Lovelace and J.~K. Townsend.
\newblock The effects of timming jitter on the performance of impulse radio.
\newblock {\em IEEE JSAC}, 20(9):1646 -- 1651, Dec. 2002.

\bibitem{dana_ubli_isit}
D.~Porrat and U.~Mitra.
\newblock {On Synchronization of Wideband Impulsive Systems in Multipath}.
\newblock In {\em Proc. IEEE ISIT 2005}, September 2005.

\bibitem{channel_uncertainty}
Dana Porrat, David Tse, and Serban Nacu.
\newblock Channel uncertainty in ultra wideband communication systems.
\newblock In Preparation, available at
  http://wireless.stanford.edu/~dporrat/ChannelUncertainty.pdf.

\bibitem{rusch_2002}
Leslie Rusch, Cliff Prettie, David Cheung, Qinghua Li, and Minnie Ho.
\newblock Characterization of {UWB} propagation from 2 to 8 {GHz} in a
  residential environment.
\newblock {\em IEEE Journal on Selected Areas in Communications}.

\bibitem{Tian}
Z.~Tian and G.B. Giannakis.
\newblock {BER} sensitivity to mis-timing in correlation-based {UWB} receivers.
\newblock {\em Proc. IEEE Globecom}, 2:441--445, Dec. 2003.
\newblock {S}an {F}rancisco, US.

\bibitem{verdu_2002}
Sergio Verd\'{u}.
\newblock Spectral efficiency in the wideband regime.
\newblock {\em IEEE Transactions on Information Theory}, 48(6):1319--1343, June
  2002.

\bibitem{VWA04}
S.~Vijayakumaran, T.~F. Wong, and S.~Aedudodla.
\newblock {On the Asymptotic Performance of Threshold-based Acquisition Systems
  in Multipath Fading Channels}.
\newblock In {\em Proc. IEEE Information Theory Workshop}, October 2004.

\bibitem{Viniotis}
Yannis Viniotis.
\newblock {\em Probability and Random Processes}.
\newblock McGraw Hill, Boston, MA, 1998.

\end{thebibliography}
\end{document}